\documentclass[aps,prl,twocolumn,showpacs,superscriptaddress,groupedaddress]{revtex4}
\usepackage{graphicx}
\usepackage{dcolumn}
\usepackage{bm}

\newcommand\br{\begin{eqnarray}}
\newcommand\er{\end{eqnarray}}
\newcommand\be{\begin{equation}}
\newcommand\ee{\end{equation}}

\newcommand\bc{\begin{center}}
\newcommand\ec{\end{center}}

\newcommand{\nn}{\nonumber \\}

\begin{document}
\title{Confining Boundary conditions from dynamical Coupling Constants}

\author
{E. I. Guendelman \footnote{e-mail: guendel@bgu.ac.il}}
\address{Physics Department, Ben Gurion University of the Negev, Beer
Sheva 84105, Israel}

\author
{R. Steiner \footnote{e-mail: roeexs@gmail.com}}
\address{Physics Department, Ben Gurion University of the Negev, Beer
Sheva 84105, Israel}

\begin{abstract}
It is shown that it is possible to consistently and gauge invariantly formulate models where the coupling constant is a non trivial function of a scalar field . In the $U(1)$ case the coupling to the gauge field  contains a term of the form
 $g(\phi)j_\mu (A^{\mu} +\partial^{\mu}B)$ where $B$ is an auxiliary field and  $j_\mu$ is the Dirac current. The scalar field $\phi$ determines the local value of the coupling of the gauge field to the 
 Dirac particle. The consistency of the equations determine the condition $\partial^{\mu}\phi j_\mu  = 0$
which implies that the Dirac current cannot have a component in the direction of the gradient of the scalar field. As a consequence, if  $\phi$ has a soliton behaviour, like defining a bubble that connects two vacuua, we obtain that the Dirac current cannot have a flux through the wall of the bubble, defining a confinement mechanism where the fermions are kept inside those bags. Consistent models with time dependent fine structure constant can be also constructed

\end{abstract}


\maketitle

\date{\today}

\maketitle
\section{Introduction}
In this paper it will be shown that an alternative coupling of gauge fields to charged particles is possible in such a way that the the coupling constants can be dynamical.  The gauge coupling has a term of the form $g(\phi)j_\mu (A^{\mu} +\partial^{\mu}B)$ where $B$ is an auxiliary field and the current $j_\mu$ is the Dirac current.
Before studying the issue of dynamical gauging, we review how the $ B $ field can be used in a gauge theory playing the role of a scalar gauge field \cite{Scalar gauge field}.That can be used to define a new type of convariante derivative.
Starting with a complex scalar field we now gauge the phase symmetry of $\phi$ by introducing a real, scalar $B( x _\mu)$ and two types of covariant derivatives as
\begin{equation}
\label{cov-ab}
D ^A _\mu = \partial _\mu + i e A_\mu ~~~;~~~ D ^B _\mu = \partial _\mu + i e \partial _\mu B ~.  
\end{equation}  
The gauge transformation of the complex scalar, vector gauge field and scalar gauge field have the 
following gauge transformation
\begin{equation} \label{gauge-trans}
\phi \rightarrow e^{i e \Lambda} \phi ~~~;~~~ A_\mu \rightarrow A_\mu + \partial _\mu \Lambda ~~~;~~~
B \rightarrow B - \Lambda ~.
\end{equation} 
It is easy to see that terms like $D ^A _\mu \phi$ and $D ^B _\mu \phi$, will be
covariant under \ref{gauge-trans} that is they transform the same way as the scalar field $ \phi $ and their complex conjugates will transfor as $\phi^*$ does. Thus one can generate kinetic energy type terms like 
$(D ^A _\mu \phi) (D ^{A \mu} \phi)^*$, $(D ^B _\mu \phi) (D ^{B \mu} \phi)^*$, $(D ^A _\mu \phi) (D ^{B \mu} \phi)^*$,
and $(D ^B _\mu \phi) (D ^{A \mu} \phi)^*$. Unlike $A_\mu$ where one can add a gauge invariant 
kinetic term involving only $A_\mu$ (i.e. $F_{\mu \nu} F^{\mu \nu}$) this is apparently not possible to do for the
scalar gauge field $B$. However note that the term $A_\mu + \partial _\mu B$ is invariant under the
gauge field transformation alone (i.e. $A_\mu \rightarrow A_\mu + \partial _\mu \Lambda$ and 
$B \rightarrow B - \Lambda$). Thus one can add a term like $(A_\mu + \partial _\mu B)(A^\mu + \partial ^\mu B)$
to the Lagrangian which is invariant with respect to the gauge field part only of the gauge transformation
in \ref{gauge-trans}. This gauge invariant term will lead to both mass-like terms for the vector gauge
field and kinetic energy-like terms for the scalar gauge field. In total a general Lagrangian which respects 
the new gauge transformation and is a generalization of the usual gauge Lagrangian, which has 
the form
\begin{eqnarray} \label{u2}
& {\cal L} = c_1 D^A _\mu \phi (D ^{A \mu} \phi ) ^* + c_2 D^B _\mu \phi (D ^{B \mu} \phi )^* 
+ c_3 D^A _\mu \phi (D ^{B \mu} \phi )^* 
\nn & + c_4 D^B _\mu \phi (D ^{A \mu} \phi )^* - V(\phi) \nonumber \\ &
- \frac{1}{4} F_{\mu \nu} F^{\mu \nu} + c_5 (A_\mu + \partial _\mu B)(A^\mu + \partial ^\mu B)~,
\end{eqnarray}
where $c_i$'s are constants that should be fixed to get a physically acceptable Lagrangian where $ c_{3}=c^{*}_{4} $ and $ c_{1}\, , c_{2}\, , c_{5} $ are real.\\
At first glance one might conclude that $B(x)$ is not a physical field, it appears that one could "gauge" it away by taking $\Lambda = B(x)$ in \ref{gauge-trans}. However in the case of symmetry breaking when one introduces a complex charged scalar field that get expectation value which is not zero, one must be careful since this would imply that the gauge transformation of the field $\phi$ would be of the form $\phi \rightarrow e^{i e B} \phi$ i.e. the phase factor would be fixed by the gauge transformation of $B(x)$. In this situation one would no longer to able to use the usual unitary gauge transformation to eliminate the Goldstone boson in the case when one has spontaneous symmetry breaking. \\ Indeed in the case when there is spontaneous symmetry breaking, the physical gauge (the  generalization of the unitary gauge) is not  the gauge $B=0$, as discussed in \cite{Scalar gauge field}, it is a gauge where the scalar gauge field $B$ has to be taken proportional to the phase of the scalar field, with a proportionality constant that depends on the expectation value of the Higgs field. Also, in general there are the three degrees of freedom of a massive vector field and the Higgs field, and therefore all together five degrees of freedom.

If there is no spontaneous symmetry breaking, fixing the gauge $B=0$  does not coincide with the gauge that allows us to display that the photon has two polarizations, this gauge being Coulomb gauge. This is true even if we do not add a gauge invariant mass term (possible given the existence of the $B$ field).
By fixing the Coulomb gauge, which will make the the photon manifestly having only two polarizations, we will have already exhausted the gauge freedom and cannot in general in addition require the gauge $B=0$. So, in Coulomb gauge where the photon will have two polarizations, the 
$B$ field and in addition the two other scalars, the real and imaginary part of $\phi$ all represent true degrees of freedom, so altogether we have five degrees of freedom, the same as the case displaying spontaneous symmetry breaking. If we add a gauge invariant mass term, even when there is no spontaneous symmetry breaking (the $c_5$ term), in the gauge $B=0$ we have three pollarizations of the massive vector field and still the real and imaginary parts of the complex scalar field $\phi$, still five degrees of freedom altogether.

Also, one can use this kind of field to define a coupling of electrodynamics to charged scalar field which enjoys only global $ U(1) $ symmetry \cite{GLOBALQED}
In "Global scalar QED" , we work with the following Lagrangian density

\begin{equation} \label{GlobalQED}
\mathcal{L}= g^{\mu\nu}\frac{\partial\psi^{*}}{\partial x^{\mu}}\frac{\partial \psi}{\partial x^{\nu}}-U(\psi^{*}\psi) -\frac{1}{4} F^{\mu \nu}F_{\mu \nu} + j_\mu (A^{\mu} +\partial^{\mu}B)
\end{equation}
 
 where
 
 \begin{equation}
 j_\mu =  ie(\psi^{*}\frac{\partial\psi}{\partial x^\mu}  -\psi\frac{\partial\psi^{*}}{\partial x^\mu})
\end{equation}

and where we have also allowed an arbitrary potential $U(\psi^{*}\psi)$ to allow for the possibility of spontaneous breaking of symmetry.
The model is separately invariant under local gauge transformations
\begin{equation} \label{GTGlobalQED}
A^\mu \rightarrow A^\mu + \partial^\mu \Lambda \textrm{;   } \ \ \ 
B \rightarrow  B - \Lambda 
\end{equation}

and the independent global phase transformations
\begin{equation}
\psi \rightarrow exp (i\chi) \psi
\end{equation} 

 The use of a gauge invariant combination $(A^{\mu} +\partial^{\mu}B)$ can be utilized for the construction of mass terms\cite{Stueckelberg}  or both mass terms and couplings to a current defined from the gradient of a scalar in the form $(A^{\mu} +\partial^{\mu}B)\partial_{\mu}A$ \cite{Guendelman}. In the non abelian case mass terms constructed along these lines have been considered by Cornwall \cite{Cornwall}.
Since the subject of this paper is electromagnetic couplings of photons and there is absolutely no evidence for a photon mass, we will disregard such type of mass terms and concentrate on the implications of the
 $(A^{\mu} +\partial^{\mu}B)j_{\mu}$ couplings.
 It is also interesting to point out the use of scalars instead of vectors fields has been studied in \cite{Singleton} in their general study of gauge procedure with gauge fields of various ranks.

\section{Confining Boundary conditions from dynamical Coupling Constants}
 In this chapter and the following one we will show that dynamical Coupling Constants can lead to confinement.
The dynamical Coupling Constants is dynamical mostly at the boundary of the confinement and out side the boundary.
 
Lets proceed with the same consideration as in the chapter before, but with Dirac field $ \psi $ and real scalar field $ \phi $, with the action:
\begin{eqnarray}\label{Dirac:boundary}
& S=\int{\bar{\psi}(i\gamma^{\mu}\partial_{\mu}-m+e\gamma^{\mu}A_{\mu})\psi \,d^{4}x} - \frac{1}{4} \int{F^{\mu\nu}F_{\mu\nu}\,d^{4}x} \nn & +\int d^{4}x [ g(\phi)\bar{\psi}\gamma^{\mu}\psi(A_{\mu}+\partial_{\mu}B)\nn & +\frac{1}{2}\partial_{\mu}\phi\partial^{\mu}\phi - V(\phi)]
\end{eqnarray}
The model is  invariant under local gauge transformations
\begin{equation} \label{GTGlobalQED2}
A^\mu \rightarrow A^\mu + \partial^\mu \Lambda \textrm{;   } \ \ \ 
B \rightarrow  B - \Lambda 
\end{equation}

\begin{equation}
\psi \rightarrow exp (ie\Lambda) \psi
\end{equation} 

The Noether current conservation law for global symmetry $ \psi \rightarrow e^{i\theta}\psi $, $\theta= constant$ is,
\begin{equation}
\partial_{\mu} j^{\mu}_{N}=(\partial_{\mu})(\frac{\partial \mathcal{L}}{\partial \psi_{,\mu}}\delta\psi) = \partial_{\mu}(\bar{\psi}\gamma^{\mu}\psi)=0
\end{equation}
The  gauge field equation, containing in the right hand side the current which  is the source of the gauge field is:
\begin{eqnarray}
\partial_{\mu}F^{\mu\nu}=(e+g(\phi))\bar{\psi}\gamma^{\nu}\psi = j^{\nu}_{Source}
\end{eqnarray}
By considering the divergence of the above equation, we obtain the additional conservation law:
\begin{eqnarray}\label{bag conservation law}
& \partial_{\mu}j^{\mu}_{Source}=\partial_{\mu}(g(\phi))\bar{\psi}\gamma^{\mu}\psi + g(\phi) \partial_{\mu}(\bar{\psi}\gamma^{\mu}\psi)\nn & = \partial_{\mu}(g(\phi))\bar{\psi}\gamma^{\mu}\psi = 0
\end{eqnarray}
If we have scalar potential $ V(\phi) $ with domain wall between two false vacuum state (see figure 2) , than because of the transition of the scalar field on the domain wall $ \partial_{\mu}(g(\phi))=\frac{\partial g(\phi)}{\partial \phi}\partial_{\mu}\phi=\frac{\partial g(\phi)}{\partial \phi}n_{\mu}f \neq 0 $ (see figure 3) 
We must conclude that $ n_{\mu}(\bar{\psi}\gamma^{\mu}\psi)\mid_{x=domain\,wall}=0 $. This means that on the domain wall there is no communication between the two sector of the domain, which give a confinement (see figure 1).

\begin{figure}[h!]
\centering
\includegraphics[scale=.4]{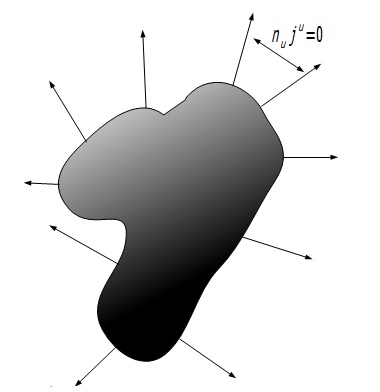} 
\caption[Caption for LOF]{{\tiny on the domain wall there is no communication between the two sector of the domain, which give a confinement}}
\end{figure}

\begin{figure}[h!]
\centering
\includegraphics[scale=.5]{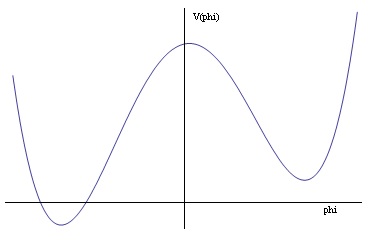}
  \caption[Caption for LOF]{{\tiny Scalar potential $ V(\phi) $ with domain wall between two false vacuum state }}
\includegraphics[scale=.5]{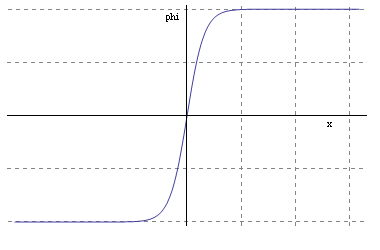} 
\caption[Caption for LOF]{{\tiny transition of the scalar field on the domain wall $ \partial_{\mu}(g(\phi))=\frac{\partial g(\phi)}{\partial \phi}\partial_{\mu}\phi=\frac{\partial g(\phi)}{\partial \phi}n_{\mu}f \neq 0 $}}
\end{figure}

\section{Confining Boundary conditions  holding at a specific region in a domain wall}
The constraint that we have gotten in the last chapter is too strong and non trivial and holds everywhere even if the gradients are small , but we wants to have constraint only in the region were the domain wall is located, so lets proceed with the same consideration as above but with coupling of the gauge field and the scalar field so the constraint will follow from an equation of motion and only at a specific location. We will see that if in  the action we add an additional term (the $ 1/l_0 $ term):
\begin{eqnarray}\label{Dirac:boundary2}
& S=\int{\bar{\psi}(i\gamma^{\mu}\partial_{\mu}-m+e\gamma^{\mu}A_{\mu})\psi \,d^{4}x} - \frac{1}{4} \int{F^{\mu\nu}F_{\mu\nu}\,d^{4}x} \nn & +\int d^{4}x [ g(\phi)\bar{\psi}\gamma^{\mu}\psi(A_{\mu}+\partial_{\mu}B)+\frac{1}{2}\partial_{\mu}\phi\partial^{\mu}\phi - V(\phi) \nn & +\frac{1}{l_{0}}\partial_{\mu} \phi (A^{\mu} +\partial^{\mu}B)]
\end{eqnarray}
The model is  invariant under local gauge transformations as in equation \ref{GTGlobalQED2}.

The Noether current conservation law for global symmetry $ \psi \rightarrow e^{i\theta}\psi $, $\theta= constant$ is,
\begin{equation}\label{Noether conservation law 2}
\partial_{\mu} j^{\mu}_{N}=(\partial_{\mu})(\frac{\partial \mathcal{L}}{\partial \psi_{,\mu}}\delta\psi) = \partial_{\mu}(\bar{\psi}\gamma^{\mu}\psi)=0
\end{equation}
The  gauge field equation, containing in the right hand side the current which  is the source of the gauge field is:
\begin{eqnarray}
\partial_{\mu}F^{\mu\nu}=(e+g(\phi))\bar{\psi}\gamma^{\nu}\psi +\frac{1}{l_{0}}\partial^{\nu}\phi = j^{\nu}_{Source}
\end{eqnarray}
We can see that we have additional long range term to the constraint \ref{bag conservation law}.
By considering the divergence of the above equation, we obtain the additional conservation law:
\begin{eqnarray}\label{bag conservation law 2}
& \partial_{\mu}j^{\mu}_{Source}=\partial_{\mu}(g(\phi))\bar{\psi}\gamma^{\mu}\psi + g(\phi) \partial_{\mu}(\bar{\psi}\gamma^{\mu}\psi) +\frac{1}{l_{0}}\partial^{\mu}\partial_{\mu}\phi \nn & = \partial_{\mu}(g(\phi))\bar{\psi}\gamma^{\mu}\psi \nn & +\frac{1}{l_{0}}\partial^{\mu}\partial_{\mu}\phi = 0
\end{eqnarray}

The variation on the action by $ \phi $ gives:
\begin{eqnarray}\label{bag EOM phi}
& \partial^{\mu}\partial_{\mu}\phi + \frac{\partial V}{\partial \phi} - \frac{1}{l_{0}} \partial_{\mu}(A^{\mu}+\partial^{\mu}B) \nn & + \frac{\partial g(\phi)}{\partial \phi}\bar{\psi}\gamma^{\mu}\psi(A_{\mu}+\partial_{\mu}B)=0
\end{eqnarray}

Lets consider a scalar potential $ V(\phi) $ with domain wall between two false vacuum state $ V(\nu_{1}) $ and $ V(\nu_{2}) $ (see figure 2) , and statically solution. Than for finite energy solution we need to demand that $ \partial_{i}\phi(\pm \infty)=0 $ and $ \phi(\infty)=\nu_{1} $ and $ \phi(-\infty)=\nu_{2} $.

From Rolle's mathematical theorem, we must conclude that at some point of the transition of the scalar field on the domain wall we have that $ \partial_{i}\partial_{i}\phi=0 $ , so equation \ref{bag conservation law 2} on some point on the transition reads:
\begin{eqnarray}
\partial_{\mu}(g(\phi))\bar{\psi}\gamma^{\mu}\psi=0
\end{eqnarray}
 because on the point of the transition were $ \partial_{i}\partial_{i}\phi=0 $, $ \partial_{i}\phi \neq 0 $ than $ \partial_{\mu}(g(\phi))=\frac{\partial g(\phi)}{\partial \phi}\partial_{\mu}\phi=\frac{\partial g(\phi)}{\partial \phi}n_{\mu}f \neq 0 $ (see figure 3) 
\\So we must conclude that $ n_{\mu}(\bar{\psi}\gamma^{\mu}\psi)\mid_{x=domain\,wall}=0 $. This means that on the domain wall there is no communication between the two sector of the domain, which give a confinement (see figure 1). Also we can see that the coupling constant far from the domain wall is constant.


\section{Consistent models with time dependent fine structure constant }
The formalism developed here provides the possibility of formulating a consistent formalism where the effective electric charge can change with space and time such possibility have been considered in cosmological contexts.
Many papers have been published on the subject of the variation of the fine structure constant.
There are some clues that show that the structure constant has been slightly variable, although this is not generally agreed.
Bekenstein \cite{Bekenstein} has shown a different approach to formulate consistently a theory with a variable coupling constant. The Oklo natural geological fission reactor has lead to a measurement that some claim it implies the structure constant has changed by a small amount of the order of $ \frac{\dot{\alpha}}{\alpha} \approx 1\times 10^{-7} $ \cite{Uzan}.
\section{Discussion and Conclusions}
It is shown that it is possible to consistently and gauge invariantly formulate models where the coupling constant is a non trivial function of a scalar field . In the $U(1)$ case the coupling to the gauge field  contains a term of the form
 $g(\phi)j_\mu (A^{\mu} +\partial^{\mu}B)$ where $B$ is an auxiliary field and  $j_\mu$ is the Dirac current. The scalar field $\phi$ determines the local value of the coupling of the gauge field to the 
 Dirac particle. The consistency of the equations determine the condition $\partial^{\mu}\phi j_\mu  = 0$
which implies that the Dirac current cannot have a component in the direction of the gradient of the scalar field. As a consequence, if  $\phi$ has a soliton behaviour, like defining a bubble that connects two vacuua, we obtain that the Dirac current cannot have a flux through the wall of the bubble, defining a confinement mechanism where the fermions are kept inside those bags. This gives rise to a condition that was considered also for example in M.I.T bag model \cite{mit bag 2} (for a review see \cite{mit bag}), but the way to obtain it is quite different.
It will interesting to study new effects that may appear when the gauge symmetry is broken. For example a modification of the M.I.T bag model when there is symmetry breaking as been study in ref \cite{Unconfined quarks and gluons}, as we pointed before the physical gauge in this case is not $ B=0 $ but it is a gauge where the $ B $ field and the phase of the Higgs field are proportional with the proportionality constant that depends on the expectation value of the Higgs field \cite{Scalar gauge field}.

The formalism developed here provides the possibility of formulating a consistent formalism where the effective electric charge can change with space and time such possibility have been considered in cosmological contexts.

\section{Acknowledgements}
We are very grateful to Douglas Singleton, Holger Nielsen, Jon Chkareuli and Rahul Kumar for very useful conversations.

\end{document}